\documentclass[universe,article,submit,oneauthors,pdftex,10pt,a4paper]{Definitions/mdpi} 
\usepackage{amsmath}
\usepackage{amsfonts}
\usepackage{amssymb}
\usepackage{graphicx}
\usepackage{epsfig}
\usepackage{mathrsfs}
\usepackage{bm}
\usepackage{microtype}
\usepackage{hyperref}
\usepackage{mathrsfs}
\usepackage{sirimacro}

\newcommand\HH{\mathscr{H}}
\renewcommand\eps\epsilon

\firstpage{1} 
\pubvolume{xx}
\issuenum{1}
\articlenumber{1}
\pubyear{2018}
\copyrightyear{2018}
\pdfoutput=1
\Title{Primordial Gravitational Waves and Reheating in a\\ New Class of Plateau-like Inflationary Potentials}
\Author{Siri Chongchitnan $^{1}$}
\AuthorNames{Siri Chongchitnan}
\address{%
$^{1}$ \quad E. A. Milne Centre for Astrophysics, University of Hull,  Cottingham Rd., Hull, HU6 7RX, United Kingdom; .chongchitnan@hull.ac.uk}
\abstract{
We study a new class of inflation model parametrized by the Hubble radius, such that $aH\propto \exp(-k\phi^n)$. These potentials are plateau-like, and reduce to the power-law potentials in the simplest case $n=2$. We investigate the range of model parameters that is  consistent with current observational constraints on the scalar spectral index and the tensor-to-scalar ratio. The amplitude of primordial gravitational waves in these models is shown to be accessible by future laser interferometers such as DECIGO. We also demonstrate how these observables are affected by the temperature and equation of state during reheating. We find that a large subset of this model can support instantaneous reheating, as well as very low reheating temperatures of order a few MeV, giving rise to interesting consequences for dark-matter production.}
\keyword{cosmology; gravitational waves; inflation)}
\begin{document}

\section{Introduction}

As cosmology progresses into the next decade, new ambitious  experiments will probe physics of the early Universe with unprecedented precision.  Some of the most exciting upcoming experiments are those that endeavour to measure the stochastic background of primordial gravitational waves, either directly using laser interferometers \cite{lisa,bbo,decigo}
or indirectly via the measurement of B-mode polarization in the cosmic microwave background (CMB) \cite{core1,core2,litebird}. A weak but measurable amplitude of gravitational waves are widely considered to be a strong evidence that an inflation-like process occurred in the early Universe.

The predicted amplitude of the inflationary gravitational-wave background depends on \ii{when} the Fourier modes corresponding to the detection frequencies exited the Hubble radius during inflation. This moment of so-called horizon-exit is normally captured by the e-fold number, $N$, defined as  the scale factor, $a(t)$, measured at the end of inflation, divided by that at cosmic time $t$:
\ba N(t)=\ln {a(t\sub{end})\over a(t)},\lab{efol}\ea
so that $N$ is initially positive and decreases to $0$ at the end of inflation.

To precisely measure $N$, nothing less than modelling the entire history of the Universe is required \cite{leach}. Key cosmic events post-inflation are now fairly well understood except for the  \ii{reheating} epoch, that is required to convert energy in the inflaton into a thermal bath, subsequently filling the Universe with radiation. Reheating is usually modelled as a post-inflationary oscillation and gradual decay of the inflaton around the minimum of the inflaton potential (see \cite{kofman,allahverdi,bassett} for reviews of reheating).

This paper demonstrates how inflationary observables from plateau-like potentials may be  affected by the details of the reheating process, thus shedding light on what we might learn about inflation from the next generation of CMB polarization and gravitational-wave experiments. We illustrate this point by constructing an interesting family of inflation models partially introduced in our previous work \cite{meinf}. These models, constructed simply by parametrizing the Hubble radius, was shown to have a remarkable relationship with the power-law potential $V(\phi)\propto\phi^k$. This work goes further by generalising the model presented in \cite{meinf}. We will show that the generalised models are a family of plateau-like potentials that are consistent with the current observational constraints from the Planck satellite \cite{planck15} for a wide range of reheating conditions.

For the rest of this work, We will work with the reduced Planck unit in which $ m\sub{Pl}/\sqrt{8\pi }=1$. We will only consider single-field inflation models with the usual canonical Lagrangian. Consequently, the inflationary Universe can be described by the Friedmann-Robertson-Walker metric with zero spatial curvature.

\section{The $\HH(\phi)$ parametrization}
In \cite{meinf}, we presented models of inflation parametrized by the inverse Hubble radius is defined as
\ba \HH(\phi) \equiv aH,\ea
where $\phi$ is the inflaton value in unit of the Planck mass, and $H\equiv \dot{a}/a$ is the Hubble parameter. We summarise the key ideas and equations of our formalism below.

The quantity $\HH(\phi)$ is a crucial link between inflationary expansion and the evolution of Fourier modes of density perturbations, since a Fourier mode with wavenumber $k$ exits the Hubble radius during inflation at the instant when $k=\HH$. It seems natural to explore the parameter space of single-field inflation models by exploiting this link.

Once $\HH(\phi)$ is specified, the inflaton potential $V(\phi)$ is completely determined using the following flowchart (all relations are exact).
\ba \HH(\phi) \quad \Longrightarrow\quad E_1=&\HH^\pr/\HH\lab{flowchart}\\
&\big\Downarrow\nn\\
 \eps  =& {2\over\bkt{E_1+\sqrt{(E_1)^2+2}}^2}\notag\\
&\big\Downarrow\nn\\
 H(\phi)=H\sub{end}&\exp\left|\int_{\phi\sub{end}}^\phi\sqrt{\eps/2}\D\phi\right|\notag \\
 &\big\Downarrow\nn\\
  V(\phi)=&H^2(3-\eps)\nn
\ea

The potential can be expressed more explicitly as
\ba V(\phi)&= H\sub{end}^2 \exp\left|\sqrt2\int_{\phi\sub{end}}^\phi\beta(\phi) \D\phi\right|(3-\beta^2),\lab{expli}\\
\text{where }\quad \beta(\phi) &\equiv \sqrt{E_1^2/2+1}-E_1/\sqrt{2}.\nn\ea

In fact, the reverse also holds: once $V(\phi)$ is specified, $\HH(\phi)$ can also be determined by solving the Hamilton-Jacobi differential equation:
\ba
\bkts{H'(\phi)}^2-{3\over2}H^2(\phi)&=-{1\over2}V(\phi),\lab{hjeq}\ea
and the definition  $\eps\equiv2\left({H'/ H}\right)^2$.

It is also useful to define the following variable
\ba E_n (\phi) &\equiv {\HH^{(n)}(\phi)\over \HH(\phi)}.\ea
The first few values of $E_n$ $(n=1,2,3)$  will determine the next-to-leading-order expressions for the inflationary observables $r$ (the tensor-to-scalar ratio) and $n_s$ (the spectral index of scalar perturbations).

By definition, inflation occurs as long as the Hubble radius shrinks, \ie
\ba\text{Inflation}\iff{\D\over \D t}\HH>0 \iff E_1>0.\lab{compi}\ea
Therefore, in our framework, an inflation model can be constructed using an \ii{increasing} function $\HH(\phi)$, with inflation ending at a maximum point. 


It is worth comparing this approach to the Hamilton-Jacobi formalism previously used to explore the parameter space for single-field inflation \cite{liddle+,lidseybig, me1,coone}. In this approach, $H(\phi)$ (the energy scale of inflation) is specified, where
\ba\text{Inflation}\iff{\D\over \D t} H<0 \iff \eps<1.\ea

Whilst $\HH(\phi)$ must be a deceasing function, $H(\phi)$ can either be decreasing or increasing\footnote{As long as $\dot{H}=H^\pr(\phi)\dot\phi<0$, which is a consequence of the Null-Energy Condition.}. The two approaches are quite different, but complementary\footnote{In particular, \cite{coone} used the Hamilton-Jacobi approach to construct a family of plateau-like potentials using truncated series with stochastic coefficients drawn from special distributions, whereas we construct similar models using a simple Gaussian function.} (see \cite{meinf, me-ee} for further dynamical comparisons).

Finally, the observables $r$ and $n_s$ can be evaluated using the usual next-to-leading order expressions \cite{lidseybig}:
\ba r &\simeq 16\epsilon[1-C(\sigma+2\epsilon)]\ff, \label{r2}\\ 
n_s&\simeq1+\sigma-(5-3C)\epsilon^2-{1\over4}(3-5C)\sigma\epsilon  +{1\over2}(3-C)\xi\nn,\ea
where  $C=4(\ln2+\gamma)-5\simeq0.0814514$ (with $\gamma$ the Euler-Mascheroni constant). The so-called `Hubble slow-roll' parameters (even though slow roll is neither required nor assumed in this work) are defined as
\ba \epsilon &\equiv 2\left({H'\over H}\right)^2,\quad \eta \equiv2{H''\over H},\quad \xi\equiv4{ H^\prime H^{\prime\prime\prime}\over
H^2}\label{flowparam},\\
\sigma&\equiv 2\eta-4\eps.\nn\ea
They are related to $E_n$ by:
\ba \eps  &= {2\over\bkt{E_1+\sqrt{(E_1)^2+2}}^2},\lab{hsr}\\
\eta &= {\eps(2E_2 + 3) -1\over 1+\eps},\nn\\
\xi &= {\eps\over (1+\eps)^3}\bigg(3\eps^3-2\sqrt2\eps^{5/2}E_3-2E_2\eps^2+8\eps (E_2)^2-3\eps^2 \ldots\notag\\
&-4\sqrt2E_3\eps^{3/2}+28\eps E_2+17\eps-2\sqrt{2}E_3\sqrt{\eps}-2E_2 -9 \bigg).\nn
\ea
Expressions \re{r2} have been shown to be in very good agreement with the numerical results calculated using the Mukhanov-Sasaki formalism even for models that do not obey the usual slow-roll conditions ($\eps_V, |\eta_V|\ll1$) \cite{mepbh}. Significant inaccuracies may occur if inflation is interrupted by a brief period of \ii{fast roll}, causing a feature in the scalar power spectrum \cite{adams,ashoorioon}. This situation does not occur in our present investigation. 



\section{Reheating and $e$-folding}

The amount of inflation is measured by the number of e-folds, $N$, defined in Eq. \ref{efol}. We will be particularly interested in the value of $N$ corresponding to the moment when CMB-scale perturbations exited the Hubble radius. This number, denoted $N_*$, is required to be around 60 to solve the horizon problem. Although it is possible to discuss inflationary predictions by simply specifying some plausible values $N_*$, the predictions for observables in some models are highly sensitive to the choice of $N_*$. In fact, it is quite easy to make the observable predictions more precise and physically meaningful by using a simple two-parameter model of reheating, which we will discuss below.

In the context of single-field inflation model, we define the corresponding inflaton value, $\phi_*$, via the relation
\ba \diff{N}{\phi}={H\over2H^\pr}\implies N_*=\int^{\phi_*}_{\phi\sub{end}}{H\over2H^\pr}\D\phi.\ea

To calculate $N_*$ (or $\phi_*$) in a given inflation model, one must also incorporate all post-inflation physics relevant to the evolution of each Fourier mode. To this end, we postulate a period of reheating between the end of inflation and the onset of radiation dominated era. In this work, we will adapt the reheating parametrization from the work of Martin and Ringeval \cite{MR,martin}, which was subsequently  used by a number of previous authors (\eg \cite{munoz, rehagen, ringeval}). In this formalism, reheating is parametrized by two variables: the temperature, $T\sub{reh}$, and the mean equation of state, $\bar{w}$, of the effective fluid during reheating. We briefly comment on the possible values of these parameters below.

The temperature is related to the energy density during reheating by
\ba \rho\sub{reh}={\pi^2\over30}g_* T\sub{reh}^4,\ea
where $g_*$ is the relativistic degree of freedom at that time. We take $g_*=100$ in this work, although our results are insensitive to the possible small variation in the theoretical value of $g_*$ \cite{kolb}. A lower limit on $T\sub{reh}$ is around a few MeV, for consistency with Big-Bang nucleosynthesis (BBN) predictions \cite{salas,choi}. An upper bound for $T\sub{reh}$ also exists, stemming from the case when the entire energy density in the inflaton decays instantaneously at the end of inflation, \iee when $\rho\sub{reh}=\rho\sub{end}$.

The mean equation of state, $\bar{w}$, arises from modelling the post-inflation plasma as a perfect fluid. A conservative bound for $\bar{w}$ is 
$$-{1\over3}<\bar{w}<1.$$
The lower bound follows from the condition that inflation ends, whilst the upper bound follows from the dominant energy condition. However, many reheating models in the literature require $\bar{w}$ in the smaller range
$$0<\bar{w}<0.25,$$
(see for example \cite{podolsky}). For the remainder of our work, we will plot results using these 4 boundary values of $\bar{w}$, namely, $-1/3$, 0, 0.25 and 1.

The two reheating parameters are encapsulated by the parameter $R\sub{rad}$ given by \ba \ln R\sub{rad}={1-3\bar{w}\over 12(1+\bar{w})}\ln{\rho\sub{reh}\over\rho\sub{end}}\lab{R1},\ea
where $\rho\sub{end}=3H^2\sub{end}$. We can then solve the following algebraic equation for $\phi_*$:
\ba&\ln R\sub{rad} =\nn\\
&N_*+ \ln\bkt{k_*\over{\rho^{1/4}_\gamma}} -{1\over2}\ln\bkt{H_*\over\sqrt3}-{1\over2\sqrt2}\left|\int_{\phi\sub{end}}^{\phi_*}\sqrt\eps \D \phi\right|.\lab{mainmain}\ea
(This equation is a variant of  Eq. 15 in \cite{MR}). Here $\rho_\gamma=3H_0^2\Omega_\gamma$ is the present energy density in radiation (we assume $\Omega_\gamma = 2.471\times10^{-5}h^{-2}$). Throughout this work we will take the pivot CMB scale to be $k_*=0.05$ Mpc$^{-1}$. At this scale, the Hubble parameter can be normalised using the next-to-leading-order formula
\ba H_*^2={8\pi^2\eps_*\mc{P_R}(k_*)\over(1-\eps_*+{C-3\over8}\sigma_*)^2}.\ea
where the value $\mc{P_R}(k_*)=2.20\times10^{-9}$ is used to normalise the power spectrum of scalar perturbations.

Once $\phi_*$ is known, the values of $r$ and $n_s$ can be calculated using Eqs. \ref{r2}.





\section{The Generalised Gaussian model}

We can now perform the calculation of reheating effects in an inflation model parametrized by
\ba\HH(\phi)\propto e^{-(\alpha\phi)^n},\quad\alpha>0.\lab{gen}\ea
We shall refer to this model as the \ii{Generalised Gaussian} (GG) model.

In this model, $n$ is an even positive integer, so that $\HH(\phi)$ is an increasing function along the branch $\phi<0$ (which is not problematic due to the even symmetry of $\HH$ and the $t\to-t$ transformation). It is also possible to extend the range of $n$ to any positive real number by performing the symmetrization $\HH\propto  \exp\bkt{-\bkt{\alpha|\phi|}^n}$. Inflation ends at the maximum point $\phi=0$.

The Gaussian case $n=2$ is remarkable because, as shown in \cite{meinf}, it gives essentially the same predictions in the $n_s$-$r$ plane as those from the well-known power-law (monomial) models $V(\phi)\propto \phi^k$, where $k$ is related to the GG model parameter $\alpha$ by:
\ba  k = {1\over2\alpha^2}.\lab{PL}\ea
In fact, this is no coincidence: the inflaton potential for $n=2$ does in fact reduce to the power-law form to a very good approximation, as we shall see shortly.

\subsection{The potential for $n=2$}

\begin{figure} 
  \centering
   \includegraphics[width=2.8in]{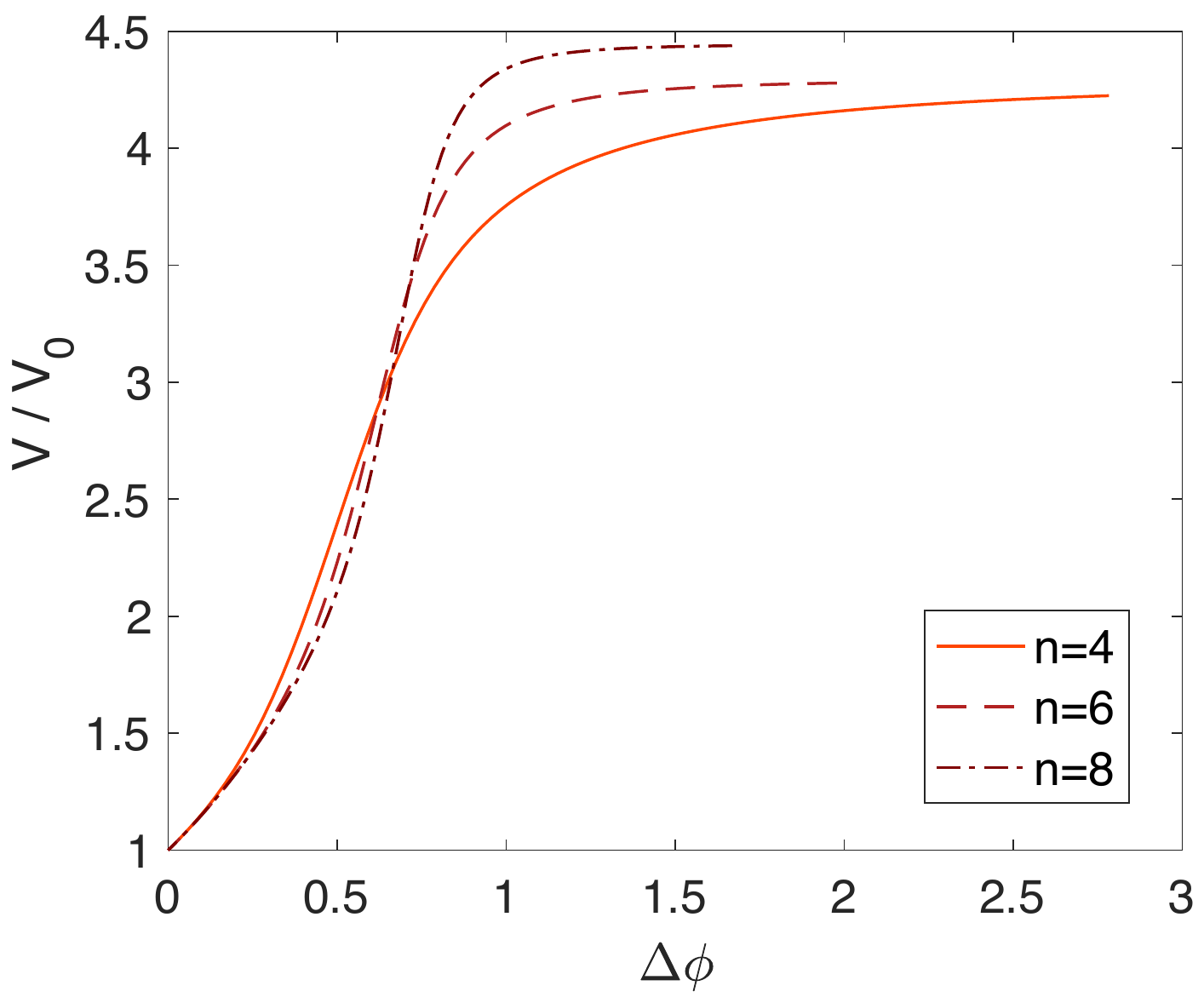} 
   \caption{The inflaton potential $V(\phi)/V(0)$ for the Generalised Gaussian model $\HH\propto\exp(-\phi^n)$, with $n=4,6,8$. On each curve,  the bottom left corner marks the end of inflation at $\phi=0$.  All these potentials are consistent with Planck's constraints in the $(n_s, r)$ plane.}
   \label{fig_potential}
\end{figure}

It is instructive to see how the potentials for the GG models look like. Fig. \ref{fig_potential} shows the potentials $V(\phi)$ for $n=4,6,8$ for a fixed value of $\alpha=1$, obtained numerically by following the algorithm \re{flowchart}. These potentials are all consistent with Planck's $2\sigma$ constraints in the $n_s$-$r$ plane. Evidently these potentials are plateaus of increasing steepness. We note the generic preference of observational data for plateau-like models \cite{planckinflation,ijjas,guth2}.

The potential for the case $n=2$ can in fact be obtained analytically. Following the flowchart, we obtain 
\ba V(\phi)&\propto (3-\beta^2)\beta^{-k}e^{k(1-\beta^2)/2},\lab{beta}\\
\text{where}\qquad \beta&=\sqrt2\alpha^2\phi+\sqrt{2\alpha^4\phi^2+1},\nn\\
k&={1\over2\alpha^2},\nn\ea
valid for $\phi\leq0$. 

This complicated potential has a very simple first-order approximation. We note that since $\phi\leq0$, $\beta$ is a small, positive number. Therefore, 
$$V(\phi)\sim \beta^{-k}$$
Furthermore, 
$$\beta^{-1}=-\sqrt{2}\alpha^2\phi+\sqrt{2\alpha^4\phi^2+1},$$
so, to lowest order in $\phi$,
\ba V(\phi)\sim \phi^k,\ea
which is simply the power-law potential. This proof strengthens the result in \cite{meinf} in which we showed that the predictions in the $n_s$-$r$ plane for the Gaussian $\HH(\phi)$ coincide what those of the power-law potentials.

\subsection{The potential for $n>2$}

For $n>2$, the potentials shown in Fig. \ref{fig_potential} can be expressed as 

\ba
V(\phi) &\propto (3-\beta^2 )\exp\left|\mc{I}\right|,\\
\text{where}\quad  \beta&\equiv -x+\sqrt{x^2+1},\lab{bex}\\
x&\equiv -{n\over\sqrt2}\alpha^n\phi^{n-1},\\
\mc{I}&\equiv\int_0^\phi\beta(\phi)\D\phi.\nn
\ea

If $n$ is an even integer, then $\beta$ is a small positive number converging to zero for a sufficiently large $n$. This suggests that $V(\phi)$ is a plateau of increasing flatness as $n$ increases.


If $n$ is large, an interesting limit can be observed if we introduce the bijective transformation $x=\sinh X$. We see from Eqs \re{bex} that 
$$\beta=-\sinh X+\cosh X = e^{-X}\approx {1\over e^X+e^{-X}}= {1\over2}\sech X,$$
in the limit that $X$ (or $x$) is large. For instance, this limit could arise when $\alpha>1$ and $n$ is large. This limit gives
\ba V(X)\propto 3-\beta^2 \approx V_0+V_1\tanh^2 X.\ea
This corresponds to the potential for the $\alpha$-attractor  `T-model', which are also plateaus of increasing flatness \cite{carrasco,kallosh}.



\begin{figure*} 
   \centering
   \includegraphics[width=\textwidth]{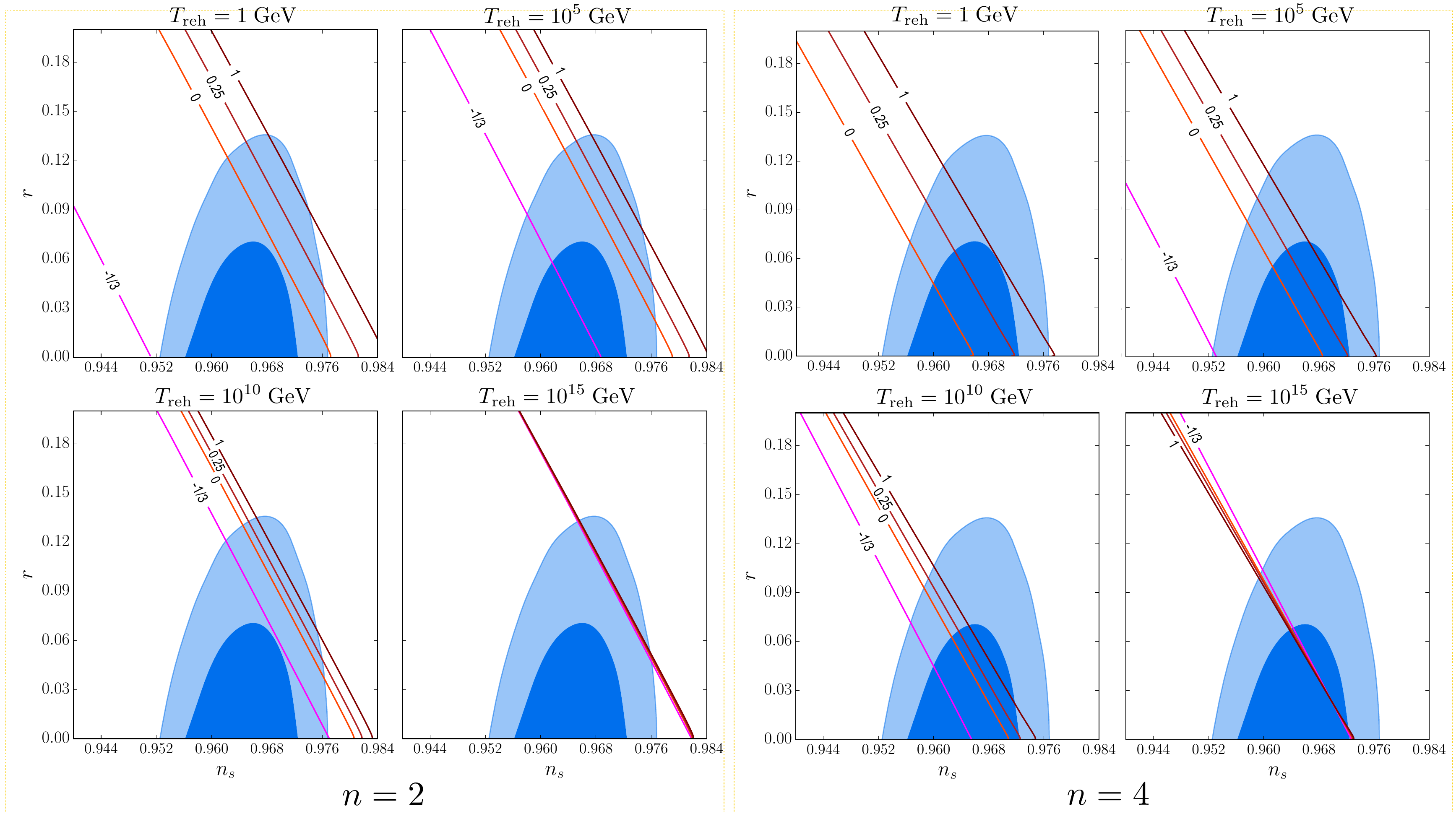}   
   \caption{Predictions in the $n_s$-$r$ plane for the Generalised Gaussian Model \re{gen} with $n=2$ (left block of four panels) and $n=4$. Each block contains four panels for reheating temperatures $T\sub{reh}=1-10^{15}$ GeV. In each panel, the lines show predictions for the reheating equation of state $\bar{w}=-1/3,0, 0.25$ and 1. On each line, as the model parameter $\alpha$ is varied from small to large, $r$ decreases steadily towards zero. See \S\ref{resu} for discussion.}
\lab{panels}
\end{figure*}
\section{Results}
\subsection{$n_s$ and $r$}\lab{resu}

Fig. \ref{panels} shows the results obtained when the GG model is analysed in the $n_s$-$r$ plane. The block of 4 panels on the left shows the predictions for $n=2$ (\iee power- law potentials), and the panels on the right are for $n=4$. Each panel corresponds to different reheating temperature, namely, $T\sub{reh}=1, 10^5, 10^{10}$ and $10^{15}$ GeV. In each panel, there are 4 lines corresponding to the values of the mean equation of state $\bar{w}=-1/3$, 0, 0.25 and 1 (from left to right, as indicated in the figure). On each line, the value of $\alpha$ is varied. As $\alpha$ increases, $r$ decreases towards zero, so that  Planck's constraints currently rule out $\alpha\ll1$.

We observe that, firstly, the predictions of the GG models in this plane tend to spread out more at lower reheating temperatures. The four lines converge at $T\sub{reh}\sim10^{15}$ GeV, where reheating is instantaneous. This is because, if reheating takes no time at all, the value of $\bar{w}$ during reheating becomes irrelevant.

We also see that increasing $n$ from 2 to 4 (or higher) displaces the lines to the left. This is in fact a generic behaviour that we observe in the GG models: higher-order GG models can be thought of as a shift in the power-law predictions towards the observationally viable region.


\subsection{Reheating temperature}

\begin{figure}[t] 
   \centering
   \includegraphics[width=2.6in]{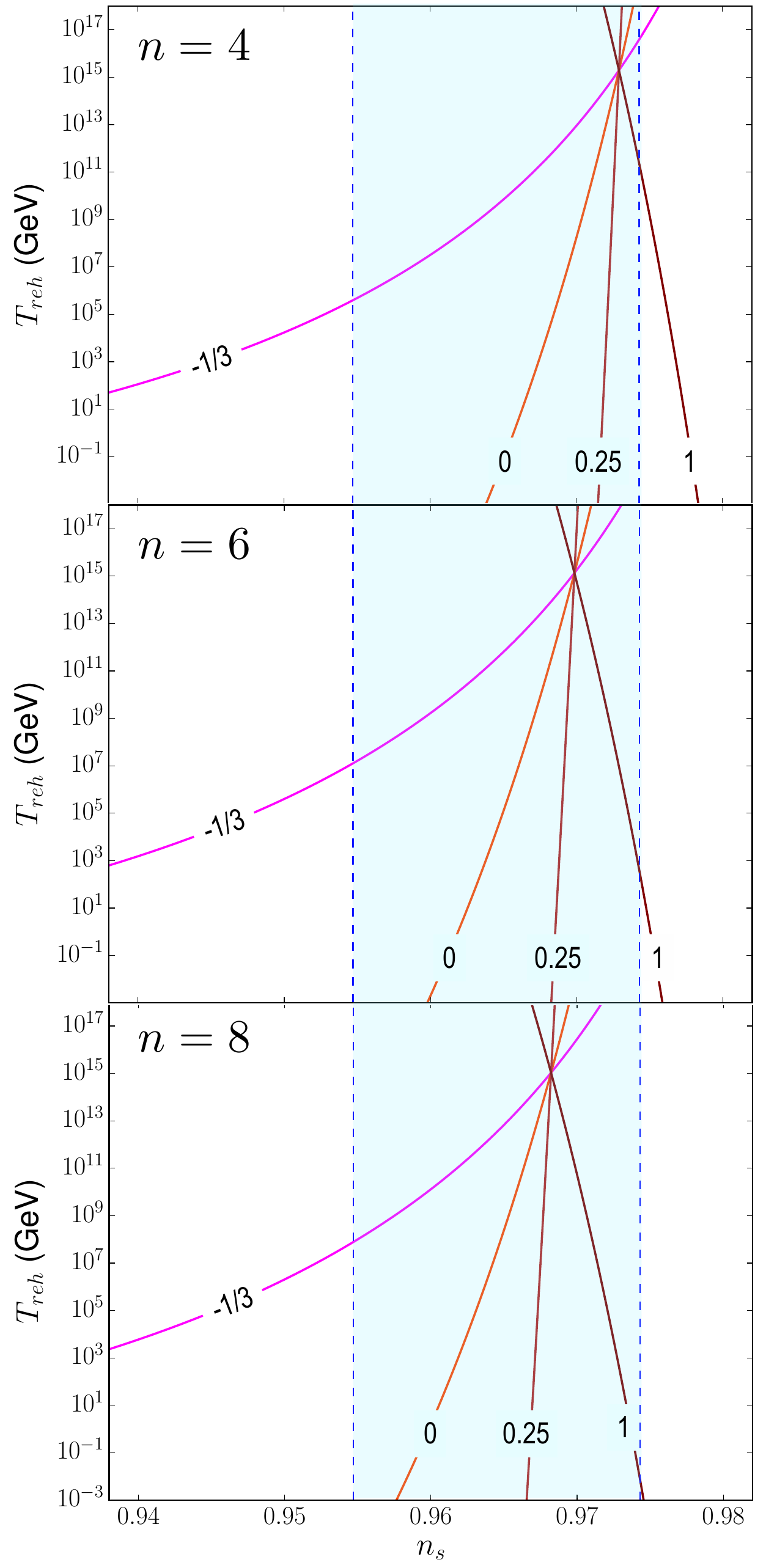} 
   \caption{Predictions in the $n_s$-$T\sub{reh}$ plane for the Generalised Gaussian Model \re{gen} with $n=4,6,8$, with $\alpha=1$. The shaded region shows Planck's $2\sigma$ constraint on $n_s$. The curves intersect where reheating would occur instantaneously.}
   \lab{triple}
\end{figure}

Fig. \ref{triple} shows the effect of varying $T\sub{reh}$ on the value of $n_s$ for the GG model with $n=4,6,8$. The physically interesting range of $T\sub{reh}$ is from  a few MeV  (the lowest reheating temperature allowed by data) to around $\sim10^{15}$ GeV where instantaneous reheating occurs, and the lines intersect as before

We note that the case $n=2$ (power law) has been previously studied in detail in \cite{rehagen,munoz}, and since we were able to reproduce their results with excellent agreement, we do not present this case here.

The shaded region is the $2\sigma$ constrain on $n_s$ from Planck ($n_s=0.9645\pm0.0098$). We chose $\alpha=1$ in all these models, which are consistent with Planck's constraints on $r$.

We observe a shift towards the left in all the curves as $n$ increases, meaning that the higher-order GG models are able to accommodate a wide range of reheating temperatures, even for $\bar{w}=1$. Furthermore, instantaneous reheating becomes observationally viable with $n>2$, even though it has been ruled out for power-law potentials (these models produce too high a value of $r$, as previously observed in \cite{rehagen}). We also note that the intersection point (corresponding to the energy scale at the end of inflation) moves slightly to lower temperatures with increasing $n$.

At the lower end of the reheating-temperature scale, we observe that higher-order GG models can  comfortably accommodate low reheating temperature of order $\sim$MeV, which will have interesting consequences for dark-matter production in the early Universe \cite{gelmini,choi}. The only exception is in the extreme case $\bar{w}=-1/3$, in which case $T\sub{reh}$ must be at least $\sim10^{6}- 10^{7}$ GeV.  




\subsection{Primordial gravitational waves at 1 Hz}

With the celebrated detections of gravitational waves from binary systems by LIGO \cite{ligo}, the hunt for gravitational waves is now progressing at a more fervid pace than ever. The most tantalising goal for the next generation of space-based laser interferometers such as BBO \cite{bbo} and DECIGO \cite{decigo} is the direct detection of a stochastic background of primordial gravitational waves, which would be a highly convincing evidence for an inflationary event in the early Universe (barring other exotic possibilities \cite{brandenberger}). These space-based interferometry have been proposed to operate in the optimal frequency window of around 0.1$-$10 Hz, in contrast with LIGO which focuses on frequencies around $100$ Hz. Unfortunately, LISA \cite{lisa} will not be sensitive to inflationary gravitational waves (at least not in the simplest scenario of canonical single field inflation). See \cite{megw, buonanno, guzzetti} for reviews of direct detection of primordial gravitational waves. 

Using the result from our previous work \cite{megw}, it is straightforward to show that the amplitude of primordial gravitational waves from inflation can be quantified by the dimensionless energy density:
\ba 
\Omega\sub{gw}(k)h^2&\approx 4.36\times10^{-15}\,r\mc{J}(k),\lab{gweq}\\
\text{where}\quad\mc{J}(k)&\equiv\exp\bkt{-\sqrt{2}\int_{\phi_*}^{\phi\sub{gw}}\sqrt{\eps}\D \phi}.\notag
\ea
The upper limit, $\phi\sub{gw}$, in the integral refers to the field value corresponding to the e-fold number when the mode with wavenumber $k\sub{gw}$ (or frequency $f= k\sub{gw}/2\pi$)  exited the Hubble radius. Given that the CMB pivot scale exited the Hubble radius at $N=N_*$, it follows that the smaller-scale mode exited the Hubble radius at
\ba N(\phi\sub{gw})= N_*-\ln\bkt{k\sub{gw}\over k_*}-\int_{\phi_*}^{\phi\sub{gw}}\sqrt{\eps\over2}\D\phi.\ea
Therefore, $\phi\sub{gw}$ can be solved numerically from the equation: 
\ba \int_{\phi\sub{gw}}^{\phi\sub{end}}{1\over\sqrt{2\eps}}\D\phi-\ln\bkt{{k\sub{gw}\over k_*}}-\int_{\phi_*}^{\phi\sub{gw}}\sqrt{\eps\over2}\D\phi=0.\lab{solvemee}\ea
$\Omega\sub{gw}h^2$, $n_s$ and $r$ can therefore be calculated on scale $k\sub{gw}$. Note that the last term in \re{solvemee} is a small correction which accounts for deviation from $H=$ constant between $\phi\sub{gw}$ and $\phi_*$. It is almost negligible in the GG model in comparison with the other terms.

\begin{figure}[t] 
   \centering
   \includegraphics[width=6in]{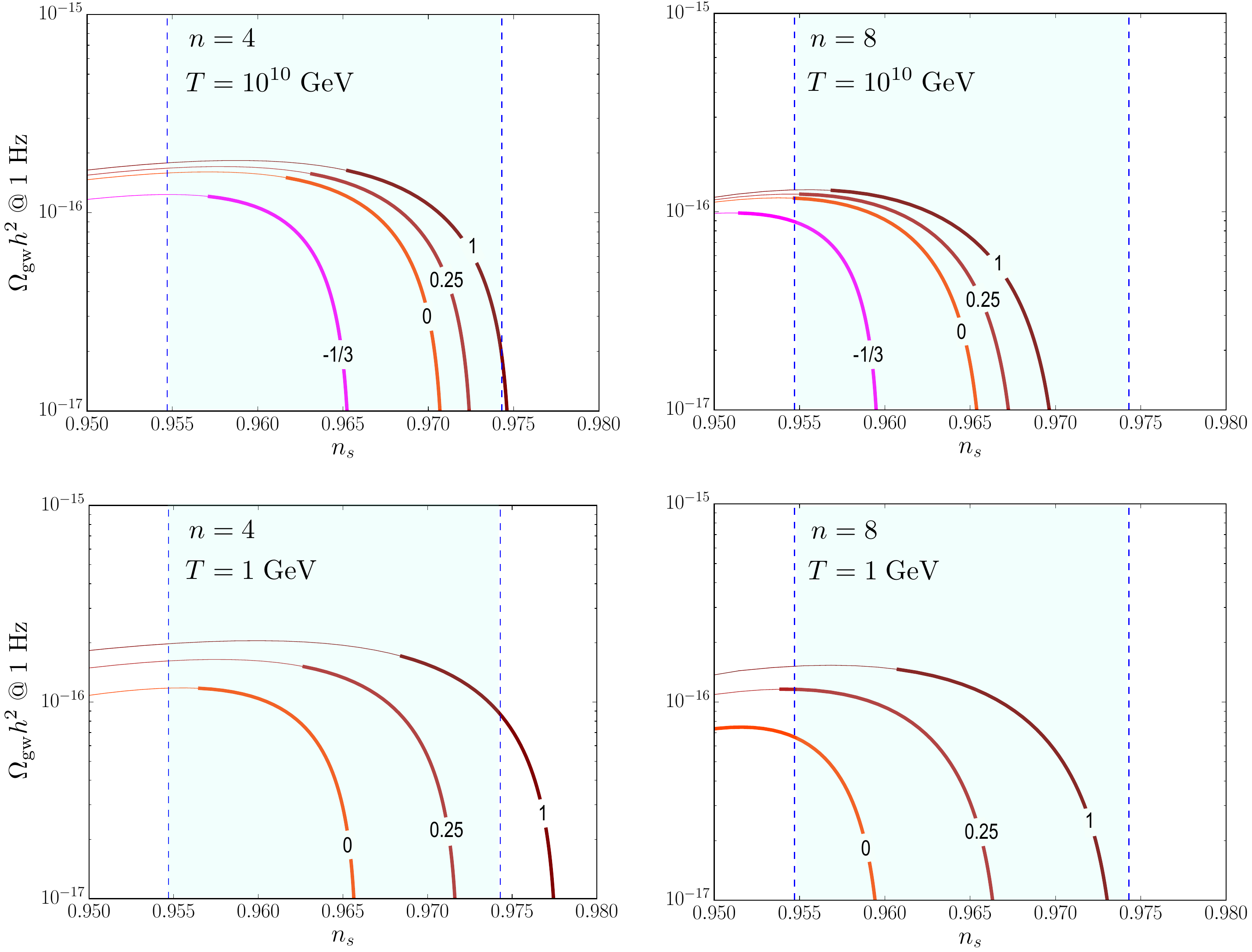} 
   \caption{The amplitude of inflationary gravitational waves, $\Omega\sub{gw}h^2$ measured at 1 Hz (Eq. \ref{gweq}) plotted against $n_s$, for the GG model with $n=4$ and 8, and $T\sub{reh}=1$ and $10^{10}$ GeV. The shaded region is the $2\sigma$ constraint on $n_s$ from Planck. Various curves in each panel correspond to the values of $\bar{w}$ as before. The thick part of each line corresponds to where $r<0.07$  \cite{bicep}. In the lower panels, we omit the $\bar{w}=-1/3$ curves as they are already ruled out by the constraint on $n_s$.}
   \label{fig_gwns}
\end{figure}

Fig. \ref{fig_gwns} shows $\Omega\sub{gw}h^2$ measured at 1 Hz  plotted against $n_s$, for the GG model with $n=4$ and 8, and $T\sub{reh}=1$ and $10^{10}$ GeV. On each line, $\alpha$ is varied, with $\Omega\sub{gw}h^2\to0$ as $\alpha$ increases. The thick portion of each line corresponds to the $\alpha$ values which give  $r<0.07$, corresponding to the BICEP+Planck joint constraint \cite{bicep}.

As before, we observe the shift of the curves to the left as $n$ increases, and the clustering of lines as $T\sub{reh}$ increases. 

Increasing the value of $\bar{w}$ also increases the gravitational wave amplitude. However, increasing $T\sub{reh}$ could either increase or decrease $\Omega\sub{gw}h^2$.

Generally, the GG models are able to produce primordial gravitational waves with amplitude as large as $\Omega\sub{gw}h^2\sim10^{-16}$. Such models are typically the least `plateau-like' and they will be the first to be ruled out by BBO/DECIGO, which, interestingly, will probe the turnover region of these curves. The steep plunge in these curves correspond to very flat plateau-like potentials. We can therefore deduce that for such potentials, an upper bound on $\Omega\sub{gw}$ and a tightened limit on $n_s$ will be effective in constraining $\bar{w}$.

\begin{figure}[t] 
   \centering
   \includegraphics[width=6.5in]{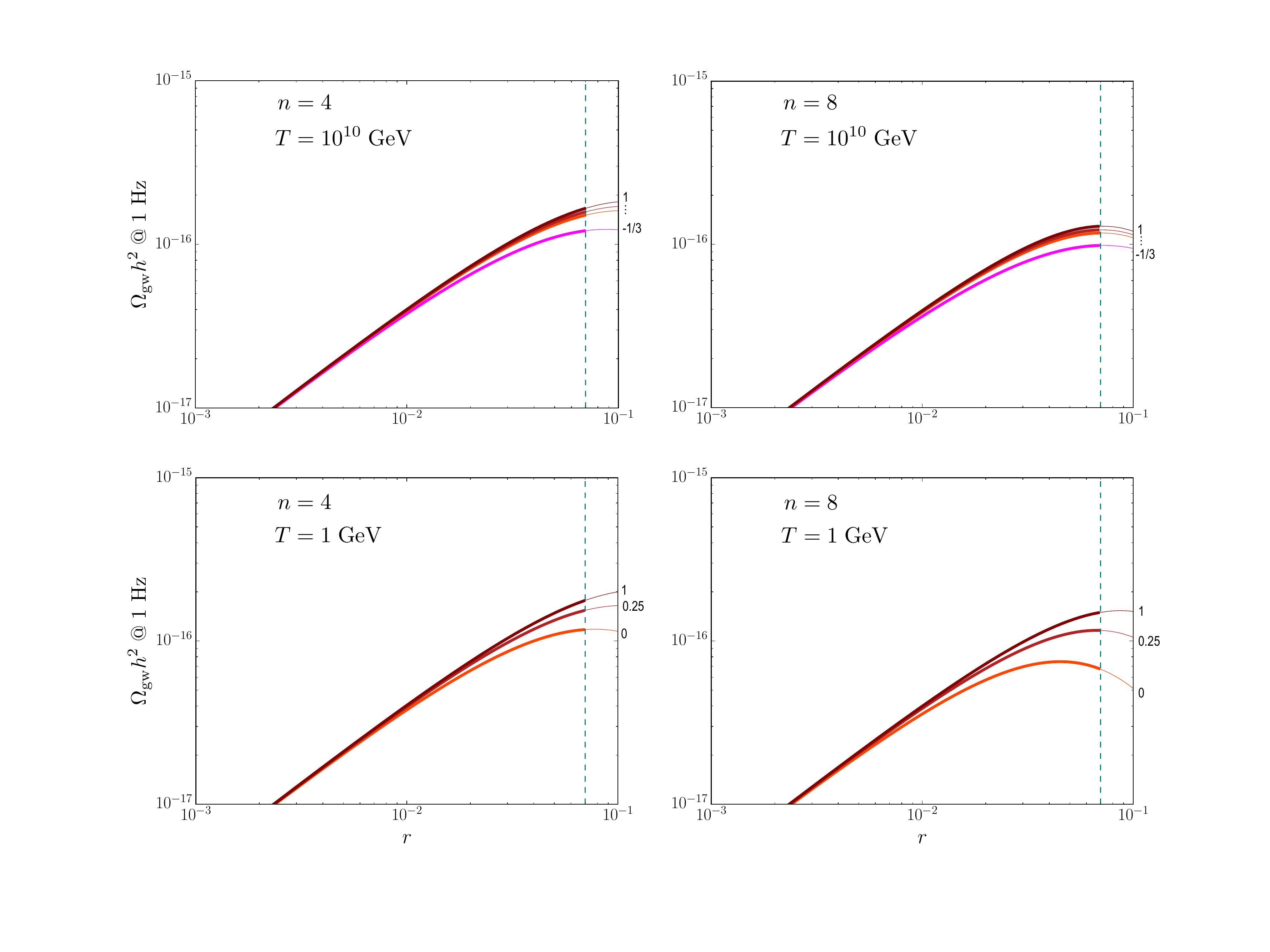} 
   \caption{$\Omega\sub{gw}h^2$ plotted against $r$ for the same models as in Fig \ref{fig_gwns}. The dashed vertical line in each panel shows the current upper bound $r<0.07$.}
   \label{fig_gwr}
\end{figure}

Fig. \ref{fig_gwr} shows $\Omega\sub{gw}h^2$ for the same models but plotted against $r$. The dashed vertical line in each panel shows the current upper bound $r<0.07$. Unlike the $n_s$-$\Omega\sub{gw}h^2$ plane, the curves in this plane are highly insensitive to changes in $\bar{w}$ and $T\sub{reh}$. Future B-mode constraints will place a stricter upper bound on $r$, and will essentially rule out any deviation from the slow-roll linear relationship $\Omega\sub{gw}\propto r$. This is because the potential for the GG models remain essentially flat between $\phi\sub{gw}$ and $\phi_*$, especially for higher values of $n$. 

In summary, the limits on $\Omega\sub{gw}h^2$ from future gravitational-wave experiments will be able to rule out the least plateau-like models in the GG family, and provide an upper bound for the equation of state $\bar{w}$. Extreme values of the reheating parameters $\bar{w}$ and $T\sub{reh}$ are also likely to be ruled out especially when combined with the constraint on $n_s$.



\section{Conclusions}

We have presented a study of a new family of plateau-like inflation model: the Generalised Gaussian model, its observational predictions and the sensitivity to the conditions during reheating.

The GG models are explicit realisations of plateau-like models preferred by observational data. They are simply constructed from modelling the evolution of the Hubble parameter $\HH\equiv aH\propto\exp(-(\alpha\phi)^n)$. We showed that the case $n=2$ corresponds to the power-law potential $V\propto\phi^k$ with $k=(2\alpha)^{-2}$. With increasing $n$, the steepness and flatness of the plateaus become enhanced.

In the observationally interesting region in the $n_s$-$r$ plane, the GG model predicts straight lines (for varying $\alpha$), just like the power-law potentials. Higher order models preserve the main features of the power-law predictions but laterally shift them into the observationally viable region. Hence, the GG models present an easy way to produce observationally-consistent inflation models, including those with large tensor amplitudes. Such models are prime candidates that will be targeted by the next generation of B-mode  experiments such as COrE \cite{core1,core2} and LiteBIRD \cite{litebird}. 

The reheating analysis in the $n_s$-$r$ plane shows that at low reheating temperatures, extremely low values of the mean equation of state $\bar{w}$ are already ruled out thanks to the tight constraint on $n_s$ (this conclusion applies to all $n$). For $\bar{w}$ in the plausible theoretically range ($0-0.25$), higher-order GG models are able to maintain reheating at a huge range of temperatures from a few MeV to $\sim10^{15}$ GeV, where reheating occurs instantaneously. We noted that instantaneous reheating is ruled for power-law potentials, but the GG model comfortably allows for this.

In addition, we calculated the amplitude of stochastic gravitational waves in the GG models and found interesting results when $\Omega\sub{gw}h^2$ is plotted against $n_s$ (assuming the direct-detection frequency of 1 Hz). Larger values of $\bar{w}$ result in larger $\Omega\sub{gw}h^2$. We also found that the curves have a characteristic turnover which will be accessible by post-LISA laser interferometers such as BBO and DECIGO. Combining gravitational waves, tensor modes and $n_s$ constraints will result in ruling out the least plateau-like potentials, whilst placing tighter limits on  reheating physics. If the inflationary potential is extremely flat, then an upper bound on $\Omega\sub{gw}$ and a tight limit on $n_s$ will be effective in constraining $\bar{w}$.

The analysis presented in this work is easy to generalise to other models of $\HH(\phi)$. It would be interesting to place constraints on the shape of $\HH(\phi)$ currently allowed by data. Other surprising correspondences between $\HH(\phi)$ and $V(\phi)$ may  emerge from our future investigation.

\bbb


\no\bb{Acknowledgements}: 
I am grateful to the organisers of COSMO'17 conference in Paris where part of this work was presented and helpful comments were received.
\reftitle{References}
\bibliographystyle{elsarticle-num}
\bibliography{reheating}




\end{document}